\documentstyle[12pt]{article}
\begin{document}
\newcommand{\bea}{\begin{eqnarray}}
\newcommand{\eea}{\end{eqnarray}}
\renewcommand{\thefootnote}{\fnsymbol{footnote}}
\newcommand{\be}{\begin{equation}}      \newcommand{\ee}{\end{equation}}
\newcommand{\st}{\scriptsize}
\newcommand{\fs}{\footnotesize}

\vspace{1.3in}
\begin{center}
\baselineskip 0.3in
{\LARGE {\bf \sc eigenfunctions of the two dimensional moshinsky-szczepaniak oscillator}}

\vspace {0.3in}
{\Large \sc nagalakshmi a. rao}

\vspace{-0.2in}
{\it Department of Physics, Government College for Women,}

\vspace{-0.2in}
{\it Kolar-563101, India}

\vspace{-0.2in}
{\it akrvce@yahoo.com}
 
\vspace{0.2in}
{\Large \sc b.a. kagali}

\vspace{-0.2in}
{\it Department of Physics, Jnanabharathi Campus, Bangalore University,}

\vspace{-0.2in}
{\it Bangalore-560056, India}

\vspace{-0.2in}
{\it bakagali@hotmail.com}
\end{center}

\vspace{0.3in}

\hspace{2.3in}
{\bf Abstract}

\indent
While the usual harmonic oscillator potential gives discrete energies in the non-relativistic case,
it does not however give genuine bound states in the relativistic case if the potential is 
treated in the usual way. In the present article, we have obtained the eigenfunctions 
of the Dirac oscillator in two spatial dimensions, adapting the prescription of Moshinsky.

\vspace{0.6in}

\vspace{0.2in}
\hspace{-0.25in}PACS NOs.: 03.65.Ge, 03.65.Pm
\newpage
\indent
The study of quantum harmonic oscillators has received considerable attention and hence 
it is of intrinsic interest to 
extend the model to the relativistic domain. Moshinsky{$^1$} formulated a novel 
prescription of introducing the interaction in the Dirac equation, which, besides the 
momentum is also linear in co-ordinates and subsequently showed that the non-relativistic 
form of the interaction reduces to one of the harmonic 
oscillator type. While in literatuer such a system is called the `Dirac Oscillator', we suggest
the name "Moshinsky-Szczepaniak Oscillator", after its proponents.

\indent
Several authors have addressed the Dirac oscillator in one space dimension.
Titchmarsh{$^2$} analysed the relativistic harmonic oscillator problem using the Green's 
function technique.  
Nogami et.al.{$^3$} have pointed out the interesting differences in the coherent states 
of the Dirac Oscillator and the 
non-relativistic harmonic oscillator. The relativistic extension of the one-dimensional 
oscillator in the Lagrangian formalism omegaas developed by Moreau{$^4$} and others. While Moshinsky{$^5$} 
has analysed the one-body relativistic oscillator using Group Theory, Villalba{$^6$} 
has dealt with the angular momentum operator in the Dirac equation. Dominguez-Adame{$^7$}
has anaylsed the one-dimensional Dirac Oscillator with a scalar interaction and shown the absence 
of Klein paradox. 

\indent
An interesting framework for discussing the Dirac Oscillator is a 2+1 space-time. 
Presently we explore the two-dimensional Dirac Oscillator and expose some special features 
not displayed by one-dimensional systems.

\vspace{-0.1in} 
The Dirac equation in two dimensions{$^8$} for a free particle would read
\bea
E{\psi}=({c\left(\alpha _{x}p_{x}+\alpha _{y}p_{y}\right)+\beta m_{0}c^{2}})\psi\ ,
\eea
with $m_{0}$ denoting the rest mass of the particle and $\alpha _{x}$, $\alpha _{y}$ and $\beta$
representing the standard Dirac matrices.

\indent 
The prescription of Moshinsky is extended to two dimensions and the interaction is introduced 
as follows
\bea
p_{x}\rightarrow p_{x}-i\beta m_{0}w x
\eea
\vspace{-0.4in} 
\bea
p_{y}\rightarrow p_{y}-i\beta m_{0}w y
\eea
The two dimensional Dirac equation may now be written as
\bea
\left[\beta E\!+\!i\hbar c\beta \alpha _{x}{\partial \over \partial
x}\!+\!\!ic\beta \alpha _{x}\beta m_{0}w x\!+\!i\hbar c\beta \alpha
_{y}{\partial \over \partial y}\!+\!ic\beta \alpha _{y}\beta m_{0}w
y\!-\!\!\beta ^{2}m_{0}c^{2}\right]\!\!\psi \!=\!0
\eea
It is convenient to introduce the following representation in terms of the Pauli matrices.
$$\alpha _{x}=\sigma _{x}=\left (\matrix{0 & 1\cr 1 & 0} \right),\ \alpha
_{y}=\sigma _{y}=\left(\matrix{0 & -i\cr i & 0} \right)\ {\rm and }\ \beta =\sigma
_{z}=\left(\matrix{1 & 0\cr 0 & -1} \right)\ . $$

Writing $\psi (x)=\left(\matrix{\psi _{1}\cr \psi _{2}} \right),$ the matrix form of Eq. (4) is
\newpage
\bea
\left(\matrix{E-m_{0}c^{2} & \!\!i\hbar c{\partial \over \partial
x}\!+\!\hbar c{\partial \over \partial y}\!-\!icm_{0}w x\!-\!m_{0}cw
y\cr\cr \!\!-\!i\hbar c{\partial \over \partial x}\!+\!\hbar c{\partial \over
\partial y}\!-\!icm_{0}w x\!+\!m_{0}cw y & -E-m_{0}c^{2}}
\!\!\right)\!\!\left(\!\matrix{\psi _{1}\cr\cr \psi _{2}}
\!\right)\!\!=\!\!\left(\!\matrix{0\cr\cr 0} \!\right)
\eea
The spinor equation may be written as a system of two first order coupled differential equations
\bea
\left(E-m_{0}c^{2}\right)\psi _{1}+\left(i\hbar c{\partial \over \partial
x}\!+\!\hbar c{\partial \over \partial y}\!-\!icm_{0}w x\!-\!m_{0}cw
y\right)\psi _{2}=0\!
\eea
\bea
\left(-i\hbar c{\partial \over \partial x}\!+\!\hbar c{\partial \over \partial
y}\!-\!icm_{0}w x\!+\!m_{0}cw y\right)\psi
_{1}-\left(E+m_{0}c^{2}\right)\psi _{2}=0\!
\eea
\noindent
{\bf Equation for $\psi_{1}$:}

The equation for $\psi_{1}$ is obtained by using Eq. (7) in Eq. (6).

\vspace{-0.1in} 
On simplification we obtain
$$\left(E^{2}-m_{0}^{2}c^{4}\right)\psi _{1}+\left\{ \hbar ^{2}c^{2}{\partial
^{2}\over \partial x^{2}}+\right.\hbar ^{2}c^{2}{\partial ^{2}\over \partial
{y}^{2}}-m_{0}^{2}c^{2}w ^{2}x^{2}-m_{0}^{2}c^{2}w ^{2}{y}^{2}$$
\vspace{-0.5in}
\bea
\ \ \ \ \ \ \ \ \ \ \ \ \ \ \ \ \ \, \ \ \ \ \ \ \ \ \ \ \ \ \ \ \  \ \ \ \ \ \left.+2\hbar c^{2}m_{0}w +2m_{0}c^{2}w L_{z}\right\} \psi _{1}=0
\eea
where $L_{z}=xp_{y}-yp_{z}$ is implied.

Using $p_{x}=-i\hbar${\Large ${\partial \over \partial x}$} and $p_{y}=-i\hbar${\Large ${\partial \over
\partial y}$}, it is straightforward to check that the above equation may be written in an elegant form as
\bea
\left[\left({p_{x}^{2}\over 2m_{0}}+{p_{y}^{2}\over 2m_{0}}\right)+{1\over
2}m_{0}w ^{2}\left(x^{2}+y^{2}\right)-\hbar w \right]\psi
_{1}=\left(w L_{z}+{E^{2}-m_{0}^{2}c^{4}\over 2m_{0}c^{2}}\right)\psi _{1}
\eea
Apparently, the first two terms are those that appear in the Hamiltonian of a non-relativistic 
2D harmonic oscillator. This justifies why the `potential' is called the Relativistic Oscillator Potential. 
The fact that Dirac particles remain bound by this interaction suggests the absence of Klein paradox.

\vspace{-0.1in} 
Equation (9) may also be written as
\bea
\left[\left({\partial ^{2}\over \partial x^{2}}+{\partial ^{2}\over \partial
{y}^{2}}\right)\!-\!{m_{0}^{2}w^{2}\over \hbar
^{2}}\!\left(x^{2}+y^{2}\right)\!+\!{2m_{0}\over \hbar ^{2}}\hbar w\right]\!\psi
_{1}\!=\!-{2m_{0}\over \hbar ^{2}}\!\left[wL_{z}+{E^{2}-m_{0}^{2}c^{4}\over
2m_{0}c^{2}}\right]\!\psi _{1}
\eea
\indent 
It is convenient to use plane polar co-ordinates $\left(\rho ,\phi \right)$
to obtain the solution of this equation.

\indent
Writing $\rho ^{2}=x^{2}+y^{2}$, and expressing $\bigtriangledown ^{2} $  as
\begin{eqnarray*}
\bigtriangledown ^{2}={1\over \rho }{\partial \over \partial \rho }\left(\rho
{\partial \over \partial \rho }\right)+{1\over \rho ^{2}}{\partial ^{2}\over
\partial \phi ^{2}}
\end{eqnarray*}
the wave function may be expressed as 
\bea
\psi _{1}=R\left(\rho \right)\Phi \left(\phi \right)=R\left(\rho
\right)e^{im\phi }
\eea
where $m=0,\pm 1,\pm 2,\pm 3,....$ is the angular momentum quantum number.

\vspace{-0.1in} 
Equation (10) takes the form
$${1\over \rho }\ {d\over d\rho }\left(\rho {dR\over d\rho }\right)-{m^{2}\over
\rho ^{2}}R\left(\rho \right)-{m_{0}^{2}{w}^{2}\rho ^{2}\over \hbar
^{2}}R\left(\rho \right)-{2m_{0\ }\hbar w\over \hbar ^{2}}R\left(\rho \right)+$$
\bea
{2m_{0}\over \hbar ^{2}}mw\hbar R\left(\rho \right)+{2m_{0}\over \hbar
^{2}}\left(E^{2}-m_{0}^{2}{c}^{4}\over 2m_{0}c^{2}\right)R\left(\rho \right)=0
\eea
Multiplying by $\rho ^{2}$, the above equation becomes
\bea
\rho ^{2}{d^{2}R\over d\rho ^{2}}+\rho {dR\over d\rho }+\left(k^{2}\rho
^{2}-m^{2}\right)R-{m_{0}^{2}w^{2}\over \hbar ^{2}}\rho ^{4}R=0.
\eea
\bea
{\rm Here,} \ \ \ \ \ \ \ \ \ \ \ \ \ \ \ k^{2}={2m_{0}\over \hbar ^{2}}\left[\left(m+1\right)\hbar
w+\left(E^{2}-m_{0}^{2}c^{4}\over 2m_{0}c^{2}\right)\right] \ \ \ \ \ \ \ \ \ \ \ \ \ \ \ \
\eea 
It is easy to check that  $k$ has the dimensions of inverse length and thus $k\rho $ is a dimensionless parameter.

In what follows we use the notation $z=${\Large ${m_{o}w\over \hbar }$}$\rho ^{2}$.  In terms of the variable $z$, 
the above equation may be written as
\bea
\ z^{2}{d^{2}R\over dz^{2}}+z{dR\over dz}+{1\over 4}\left\{
k_{1}z-m^{2}-z^{2}\right\} R\left(z\right)=0\ 
\eea
where
\vspace{-0.2in} 
\bea
k_{1}=k^{2}{\hbar \over
m_{0}w}=2\left(m+1\right)+{\left(E^{2}-m_{0}^{2}c^{4}\right)\over
m_{0}c^{2}\hbar w}.
\eea
While the first term refers to the oscillator part, the second terms refers to the kinetic energy of the particle.  
We try solutions of the form
\bea
R\left(z\right)=e^{-z\over 2}z^{m\over 2}\phi _{1}\left(z\right).
\eea
The unknown function $\phi _{1}\left(z\right)$ should be a constant for $z\rightarrow 0$ and
should guarantee normalisation.
Equation (15) may be written in the standard form as
\bea
z{d^{2}\phi _{1}\over dz^{2}}+\left(b-z\right){d\phi _{1}\over dz}-a\phi_{1}=0
\eea
with $b=m+1$ and $a=$ \ {\Large ${1\over 2}$}{\Large $($}$m+1-${\Large ${k_{1}\over 2})$}.
We identify this differential equation as the Kummer's equation{$^9$} whose solution can be
expressed in terms of the confluent hypergeometeric functions.  
The only admissible solution is $M\left(a,b,z\right)$. The other linearly independent solution  
$U\left(a,b,z\right)${$^{10}$} is rejected since it is irregular at infinity.

\vspace{-0.1in}
We now write the solution as 
\bea
R\left(\rho \right)=e^{{-m_{0}w\over 2\hbar }\rho ^{2}}\left({m_{0}w\over \hbar
}\rho ^{2}\right)^{m\over 2}M\left({1\over 2}\,(m+1-{k_{1}\over
2}),m+1,{m_{0}w\over \hbar }\rho ^{2}\right)
\eea
The complete wavefunction is written as 
\vspace{0.1in}
\bea
\psi _{1}=Ae^{im\phi }e^{-{m_{0}w\over 2\hbar }\rho
^{2}}\left(\left(m_{0}w\over \hbar \right)^{1\over 2}\rho
\right)^{m}M\left({1\over 2}\,(m+1-{k_{1}\over 2}),m+1,{m_{0}w\over \hbar }\rho
^{2}\right)
\eea
where $A$ is the normalisation constant. Further, as is well-known, when $a = -n$, 
the hypergeometric series terminates and defines a finite polynomial of $n^{th}$ degree.
Hence
\vspace{0.1in}
\bea
\psi _{nm}^{\left(1\right)\mathstrut }=Ae^{im\phi }e^{-{m_{0}w\over 2\hbar
}\rho ^{2}}\left(\left(m_{0}w\over \hbar \right)^{1\over 2}\rho
\right)^{m}M\left(-n-1,m+1,{m_{0}w\over \hbar }\rho ^{2}\right)
\eea
The non-negative integer $n$, which arises from the boundary condition that the wave function 
vanishes as $\rho \rightarrow \infty $, also quantizes the energy eigenvalues.

\noindent
{\bf Equation for $\psi_{2}$}

The equation for the small component of the Dirac wave function is obtained in an analogous manner.
Eliminating $\psi_{1}$ in Eq. (7) first, and then going through similar steps as before, we obtain
\vspace{0.1in}
\bea
\left[\left({p_{x}^{2}\over 2m_{0}}+{p_{y}^{2}\over 2m_{0}}\right)+{1\over
2}m_{0}w^{2}\left(x^{2}+y^{2}\right)+\hbar w\right]\psi
_{2}=\left[wL_{z}+{E^{2}-m_{0}^{2}c^{4}\over 2m_{0}c^{2}}\right]\psi _{2}
\eea
\noindent
which resembles Eq. (9).  Adapting the same procedure as before it is seen 
that the eigenfunctions can be expressed as 
\vspace{0.1in}
\bea
\psi _{nm}^{\left(2\right)}=Ae^{im\phi }e^{-{m_{0}w\over 2\hbar }\rho
^{2}}\left(\left(m_{0}w\over \hbar \right)^{1\over 2}\rho
\right)^{m}M\left(-n,m+1,{m_{0}w\over \hbar }\rho ^{2}\right).
\eea

\noindent
{\large \bf The Energy Spectrum}

The quantisation of energy demands the vanishing of the eigenfunctions at infinity, which comes from the 
fact that the first argument of $M(a,b,z)$ be equal to a negative integer or zero{$^8$}. We obtain from Eq. (20)
\bea
{1\over 2}\left(m+1-{k_{1}\over 2}\right)=-n-1
\eea
or equivalently
\bea
E^{2}-m_{0}^{2}c^{4}=4\left(n+1\right)m_{0}c^{2}\hbar w.
\eea
As is expected for central potentials, the energy eigenvalues are independent of the quantum number $m$, 
as a consequence of rotational symmetry.

\noindent
{\large \bf Non-relativistic limit}

The non-relativistic limit is obtained by setting $E=m_{0}c^{2}+{\epsilon }_{nr}$ and considering \linebreak$\epsilon _{nr}<<m_{0}c^{2}$.
It is straightforward to check that Eq. (25) may be written as
\bea
E=m_{0}c^{2}\left[1+{4\left(n+1\right)\hbar w\over m_{0}c^{2}}\right]^{1\over
2}.
\eea
Taylor expansion of the above would give us
\bea
E\approx m_{0}c^{2}+2\left(n+1\right)\hbar w-{2\left(n+1\right)^{2}\hbar
^{2}w^{2}\over m_{0}c^{2}}.
\eea
It is thus seen that the first term corresponds to the rest energy of the particle, the second term refers to the non relativistic harmonic 
oscillator energy and the third is the relativistic correction term.

\noindent
{\large \bf Results and Discussion}

The harmonic oscillator problem in quantum mechanics has far-reaching consequences.  
While the non-relativistic harmonic oscillator problem is addressed by solving the Schrodinger equation directly 
or by the 
well-known matrix formulation method or the operator method, the relativistic harmonic oscillator requires an 
altogether 
different formalism. The prescription of Moshinsky which describes the one-dimensional Dirac oscillator may well be extended to
two space dimension. It is seen that the eigenfunctions are expressed in terms of regular confluent hypergeometric functions, 
special cases of which are the well known Hermite polynomials. More importantly the eigenenergies have the appropriate non-relativistic limit.
The Dirac oscillator is thus a relativistic generalisation of the quantum harmonic oscillator. 

\indent
Unlike  the non-relativistic oscillator, 
where the energy levels
are discrete and equispaced, the relativistic oscillator no doubt has  discrete energies, 
but unevenly spaced levels. Dirac oscillator has potential applications in models of quark 
confinement in particle physics.

\noindent
{\large \bf Acknowledgements}

One of the authors (N.A.Rao) is grateful to University Grants Commission and the Department of 
Collegiate Education in Karnataka for the award of Teacher Fellowship under the Faculty Improvement Programme.
The encouragement of Prof.V.Kamalamma, Joint Director of Collegiate Education is acknowledged with gratitude.
\newpage
\baselineskip 0.2in
\hspace{2.3in}
{\Large \bf References}
\begin {enumerate}
\item M. Moshinsky and Y. F. Simon, {\it The Harmonic Oscillator in Modern Physics}, Harwood Academic Publishers, Vol.9 in 
Contemporary Concepts in Physics Series, (1996), Chapter-XI.

\item E. C. Titchmarsh {\it On the relation between the eigenvalues in relativistic and non-relativistic quantum mechanics}.
Proc. Roy. Soc. Ser. {\bf A 266}, 33-46 (1962) Quart. J. Math. 15 (1964), 193-207.

\item Y. Nogami and F.M. Toyama, {\it Coherent states of the Dirac Oscillator}, Can. J. Phys., {\bf 74} (1996), 114-121.

\item W. Moreau, Richard Easther and Richard Neutze, {\it Relativistic (an) harmonic oscillator}, Am. J. Phys., {\bf 62}(6)(1994), 531-535.

\item M.Moshinsky and Szczepaniak, {\it The Dirac Oscillator}, J. Phys. A. Math. {\bf 22} (1989), L817-819.

\item V. M. Villalba, {\it The angular momentum operator in the Dirac equation}, Eu. J. Phys. {\bf 15} (1994), 191-196.

\item F. Dominguez-Adame, {\it A relativistic interaction without Klein Paradox}, Phys. Lett. {\bf A 162} (1992), 18-20.

\item P. Strange, {\it Relativistic Quantum Mechanics}, (Cambridge University Press, 1998), Chapter-9, Sec. 2, p 269-280.

\item M. Abramowitz and I. Stegun {\it Handbook of Mathematical functions and Formulas, Graphs and Mathematical Tables} (Dover, New York, 1964).

\item Y. S. Gradshteyn and I. M. Ryzhik {\it Table of Integrals, Series and Products} (Academeic Press, New York, 1965)
\end{enumerate}

\end{document}